%% file: article.tex
\documentclass[submission,copyright,creativecommons]{eptcs}
\usepackage[english]{babel}
\usepackage{hyperref}
\usepackage{graphicx}
\usepackage{floatflt}
\usepackage{amssymb}
\usepackage{amsfonts}
\usepackage{amsmath}
\usepackage{latexsym}
\usepackage{verbatim}
\usepackage{xcolor}
\usepackage{color}
\usepackage{multirow}
\usepackage{tabularx}
\usepackage{url}
\usepackage{booktabs}
\usepackage{listings}
\usepackage{xspace}
\usepackage{enumitem}
\usepackage{groove2tikz}
\usepackage[skip=5pt]{caption}
\title{Graph- versus Vector-Based Analysis of a Consensus Protocol%
}
\author{
  Giorgio Delzanno
  \institute{DIBRIS\\Universit\`a di Genova (Italy)}
\and
  Arend Rensink
  \institute{Department of Computer Science\\University of Twente (Netherlands)} 
\and
  Riccardo Traverso
  \institute{DIBRIS\\Universit\`a di Genova (Italy)}
  \institute{FBK-irst\\Trento (Italy)}
}
\def\titlerunning{Graph- versus Vector-Based Analysis}
\def\authorrunning{G. Delzanno, A. Rensink, and R. Traverso}
\begin{document}
\input macros
\maketitle
\input introduction

\input paxos_pseudo_code
\input spec_groove
\input spec_spin
\input analysis

\input conclusions
\bibliographystyle{eptcs}
\bibliography{biblio}
\newpage
\appendix
\end{document}

%% file: macros.tex
\def\topfraction{.90}
\def\bottomfraction{.90}
\def\textfraction{.10}
\def\floatpagefraction{.90}

\renewcommand{\paragraph}[1]{\medskip\noindent\textbf{#1} }

\def\figureTextSize{\small}

\newtheorem{definition}{Definition}

\newenvironment{proof}[1][Proof]{\begin{trivlist}
\item[\hskip \labelsep {\bfseries #1}]}{\end{trivlist}}
\newenvironment{example}[1][Example]{\begin{trivlist}
\item[\hskip \labelsep {\bfseries #1}]}{\end{trivlist}}
\newenvironment{remark}[1][Remark]{\begin{trivlist}
\item[\hskip \labelsep {\bfseries #1}]}{\end{trivlist}}

\newcommand{\qed}{\nobreak \ifvmode \relax \else
      \ifdim\lastskip<1.5em \hskip-\lastskip
      \hskip1.5em plus0em minus0.5em \fi \nobreak
      \vrule height0.75em width0.5em depth0.25em\fi}

\newcommand{\tuple}[1]{\langle{#1}\rangle}
\newcommand{\Graphs}{{\cal G}}
\newcommand{\specScale}{.47}
\newcommand{\specScaleSmall}{.44}
\newcommand{\remove}[1]{}

%
%
\lstset{
    columns=flexible,
    basicstyle=\sffamily,
    showspaces=false,          
    showstringspaces=false,    
    showtabs=false,            
}

\lstdefinelanguage{groove-control}{
    morekeywords={
        package,import,function,recipe,
        alap,while,until,do,
        if,else,try,or,choice,
        true,false,any,other,out,
        node,bool,string,int,real
    },
    sensitive=true,
    morecomment=[l]{//},
    morecomment=[s]{/*}{*/},
    morestring=[b]",
    mathescape=true,
}

\lstdefinelanguage{promela}{
     morekeywords={
         active,proctype,short,byte,int,
         do,od,break,if,else,fi,
         atomic,d_step,init,of,
         inline,define,chan,run
     },
     sensitive=true,
     morecomment=[l]{//},
     morecomment=[s]{/*}{*/},
     morestring=[b]",
     mathescape=true,
}

\lstnewenvironment{groove-control}[1][]
    {\lstset{language=groove-control,#1}} 
    {}
\lstnewenvironment{promela}[1][]
    {\lstset{language=Promela,#1}}
    {}

\newcommand{\inputpromela}[1]{%
  \unsetsnippet%
  \lstinputlisting[language=promela,linerange=#1]%
                  {promela/paxos-4.pml}
  \setpromela
}

\newcommand{\setpromela}{%
  \lstMakeShortInline[language=promela]\+%
}
\newcommand{\unsetsnippet}{%
  \lstDeleteShortInline\+%
}

\newcommand{\inputgroove}[1]{%
  \unsetsnippet
  \begin{tabular}{@{}c@{}}
  \lstinputlisting[language=groove-control,linerange=#1]
                  {grammars/paxos-2.gps/control.gcp}
  \end{tabular}
  \setgroove
}

\newcommand{\setgroove}{%
  \lstMakeShortInline[language=groove-control]\+%
}

\newcommand{\inputtikz}[1]{%
  \unsetsnippet
  \input{immagini/#1.tikz}%
  \setgroove
}
\newcommand{\inputtikzrule}[1]{%
  \hfill
  \begin{tabular}[b]{@{}c@{}}
  \unsetsnippet
  \input{immagini/#1.tikz}%
  \setgroove
 \\
  Rule~\lstinline[language=groove-control]{#1}
  \end{tabular}
  \hfill
}
\newcommand{\inputctikzrule}[1]{%
  \hfill
  \inputtikzrule{#1}
  \hfill
}

\newcommand{\norm}[1]{%
  \unsetsnippet%
  #1%
  \setgroove%
}

\newcounter{pseudocode}
\newenvironment{pseudocode}[2][\hline]{%
  \stepcounter{pseudocode}%
  \def\finalise{#1}%
  \begin{tabular}[t]{@{}p{7.5cm}@{}}%
  \hline\hline
  {\bf #2} \\
  \hline
  \begin{minipage}{7.5cm}
  \begin{tabbing}
  {\bf init} \= \qquad \= \qquad\qquad \= \kill
}{%
  \end{tabbing}
  \end{minipage} \\
  \finalise%
  \end{tabular}
}
\newcommand{\pcomm}[1]{\emph{/* #1 */}}

\newcommand{\Propose}{\mathit{Propose}}
\newcommand{\Prepare}{\mathit{Prepare}}
\newcommand{\Accept}{\mathit{Accept}}
\newcommand{\Promise}{\mathit{Promise}}
\newcommand{\Learn}{\mathit{Learn}}
\newcommand{\pickNextRound}{\mathit{pickNextRound}}
\newcommand{\pick}{\mathit{pick}}
\newcommand{\rnd}{\mathit{rnd}}
\newcommand{\crnd}{\mathit{crnd}}
\newcommand{\myval}{\mathit{myval}}
\newcommand{\prnd}{\mathit{prnd}}
\newcommand{\pval}{\mathit{pval}}
\newcommand{\aval}{\mathit{aval}}
\newcommand{\lval}{\mathit{lval}}
\newcommand{\valu}{\mathit{val}}
\newcommand{\maj}{\mathit{maj}}

\newcommand{\NAC}{\mathit{NAC}}
\newcommand{\GROOVE}{\textsc{groove}\xspace}
\newcommand{\SPIN}{\textsc{spin}\xspace}
\newcommand{\PROMELA}{\textsc{promela}\xspace}

%% file: introduction.tex
\begin{abstract}
  The Paxos distributed consensus algorithm is a challenging case-study for
  standard, vector-based model checking techniques. Due to asynchronous
  communication, exhaustive analysis may generate very large state spaces
  already for small model instances. In this paper, we show the advantages of
  graph transformation as an alternative modelling technique. We model Paxos
  in a rich declarative transformation language, featuring (among other
  things) nested quantifiers, and we validate our model using the \GROOVE
  model checker, a graph-based tool that exploits isomorphism as a natural way
  to prune the state space via symmetry reductions. We compare the results
  with those obtained by the standard model checker \SPIN on the basis of a
  vector-based encoding of the algorithm.
\end{abstract}
\section{Introduction}

Automated validation of distributed algorithms like routing and consensus
protocols is a challenging task for state-of-the-art model checkers
\cite{AWN12,FHM07,SRS09,SWJ08,KVW12,DT13,bokor2010efficient}. These protocols often depend on the current network topology and
operate under asynchronous communication assumptions.  These two features are
a frequent cause of state space explosion.  To attack this problem, in
\cite{DT13} we have proposed to apply an unconventional model checking
approach based on Graph Transformation Systems (GTS).  Graph grammars provide
a declarative language to specify updates (of structure and labels) in a
graph-based representation of a dynamic system. Apart from providing an alternative description of the problem that may in itself be interesting, the formalisation of states as graphs rather than the more conventional vectors of data values opens the way towards additional techniques for state space reduction.

In this paper we focus our attention on a graph-based declarative specification
of the Paxos distributed consensus algorithm. The consensus problem requires agreement among a number of agents for a single data value. Some of the agents may fail, so consensus protocols must be fault tolerant. Initially, each agent proposes a value to all other ones. Agents can then exchange information. A correct protocol must ensure that when a node takes the final choice, the chosen value is the same for all correct agents. It is assumed here that messages can be delayed arbitrarily. A subset of processes can crash anytime and restore their state (keeping their local information) after  an arbitrary delay.

Fisher, Lynch and Patterson have shown that, under the these assumptions,
deterministically solving consensus is impossible \cite{FLP85}. In \cite{Paxos} Lamport proposed
a (possibly non-terminating) algorithm, called Paxos, addressing this problem.
Paxos is based on the metaphor of a part-time parliament, in which part-time
legislators need to keep consistent records of their passing laws. Because the
description proved hard to understand, Lamport later provided a simpler
description of the protocol in \cite{PaxosMadeSimple}. This is the version on
which we base the models in this paper.

Our declarative specification is based on graph transformation rules with symbolic conditions on node attributes, negative application
conditions and node quantification, as provided in the \GROOVE framework
\cite{GROOVE}. Salient features of the specification are:
\begin{itemize}[noitemsep]
\item We use node occurrences as abstractions of proposed values and process
  identifiers. This causes symmetries to show up as graph isomorphism between states.

\item We use a Linda-like model for asynchronous communication, in which
  message broadcasts are represented by special nodes linked to their senders,
  without associated buffers or channels. This avoids state differences due to
  irrelevant message orderings.

\item Our rules encapsulate a lot of functionality within a single (atomic) transformation step, and thus avoid intermediate states during evaluation.
\end{itemize}
We compare the resulting models computed by \GROOVE to those generated by \SPIN from a specification of the same protocol in \PROMELA. The comparison shows that the choices listed above manage to keep the graph-based state space size to a fraction of that of a more traditional vector-based specification, enabling the analysis of larger problem instances despite the inherent complexity of graph transformation.


Our method can be seen
as an attempt of combining declarative reasoning and efficient search methods
for this class of protocols. Furthermore, it represents an alternative to
standard model checking frameworks based on (unstructured) symbolic
representations, e.g., BDDs.

%% file: paxos_pseudo_code.tex
\section{The Paxos Consensus Algorithm}\label{paxos}

The description of Paxos in
\cite{PaxosMadeSimple} distinguishes three separate agent roles:
\emph{proposers} that can propose values for consensus, \emph{acceptors} that
accept a value among those proposed, and \emph{learners} that learn the
accepted values and eventually choose one of them. We present the protocol on the basis of a pseudo-code description from the lecture notes \cite{MMH13}.

\begin{figure}[t]
\figureTextSize
\begin{tabular}{p{7.5cm}p{7.5cm}}
\input{proposer} 
&
\input{acceptor}\hfill
\input{learner}\hfill
\end{tabular}
\caption{Pseudo-code of the Paxos protocol}
\label{pseudocode}
\end{figure}

In a first step, the proposer selects a fresh round identifier and broadcasts it to all acceptors, in a message called $\Prepare$.  It then collects votes for that round number from live acceptors.
Acceptors' replies, called $\Promise$s, contain the round number and a pair consisting of the last round and value that they promised in previous rounds (with the same or a different proposer).
Rounds and values are initialized to the default of $-1$, and only change  upon $\Accept$ messages.
When the proposer checks that a majority is reached, it selects a value to submit again to the acceptors.  For the selection of this value, the proposer inspects every $\Promise$ received in the current round and selects the value with the highest non-default round; if it did not receive a non-default value, it uses its own initial proposal ($\myval$). It then submits the current round and the chosen value to the acceptors, in a message called $\Accept$.

Acceptors wait for proposals (i.e., $\Prepare$ messages) of round identifiers but consider only those that are \emph{fresh}, in the sense of being higher than the last one they have seen so far. If the received round is fresh, acceptors answer with a $\Promise$ not to accept proposals with smaller round numbers. Since messages might arrive out-of-order, even different $\Prepare$s with increasing rounds of the same proposer might arrive in arbitrary order (this justifies the need of the $\Promise$ message). Acceptors also wait for $\Accept$ messages: in that case local information about the current round is updated and, if the round is fresh, the accepted pair $(\rnd,\aval)$ is forwarded to the learner, in a message called $\Learn$.

A learner collects votes ($\Accept$ messages) on pairs $(\rnd,\lval)$ sent by acceptors and waits to detect a majority for one of them. When a majority is detected, the $\lval$ component is chosen.

\medskip\noindent
The pseudo-code of the algorithm, based on 
\cite{MMH13}, is given in Fig.\ \ref{pseudocode}. Majority is defined as $\maj$. The pseudo-code includes a special $\Propose$ message that
corresponds to an external command sent to the node in order to inject a new
proposal (a value) into the whole system. In the proposer code, $\pickNextRound$ must return a fresh
value (w.r.t.\ all processes) for the next round, and $\pick$
must return the value associated to a tuple with highest round number. 
We use $\oplus$ to denote multiset union (which is required to count multiple occurrences of the same pair).

The protocol is guaranteed to reach consensus for $\maj\geq
\lceil(\#A+1)/2\rceil$, where $\#A$ denotes the size of the set $A$ of correct
acceptors, and only if acceptors and learners have enough time to take a
decision (i.e., to detect a majority).  If proposers indefinitely inject new
proposals, the protocol may diverge. In this paper we will concentrate on the
following limited notion of correctness:
\begin{definition}[safety]\label{safety}
The protocol is correct if, when a value is chosen by a learner, it has been proposed by a proposer, and no other value has been chosen by any learner in previous rounds of the protocol.
\end{definition}
This means that, whenever a value is chosen by a learner, any successive
choices always select the same value (possibly with larger round identifiers);
i.e., the algorithms stabilizes w.r.t.\ the value components of tuples sent to
the learners.

\paragraph{Simplifying assumptions.}
In both the \GROOVE and the \SPIN model presented in this paper, we have made the following simplifying assumptions about the protocol:
\begin{itemize}[noitemsep]
\item
Proposers never send more than one $\Prepare$ message. This does not restrict the protocol for the purpose of the correctness criterion in Def.~\ref{safety} (the effect of multiple $\Prepare$ messages may be mimicked by increasing the number of proposers) but it causes the protocol to always terminate.

\item There is only a single learner. This cannot affect the correctness of the protocol either, as all learners have access to exactly the same information and hence are bound to have the same behaviour. In other words, any error in a scenario with multiple learners must necessarily occur already with a single learner.
\end{itemize}

%% file: proposer.tex
\begin{pseudocode}{{Paxos --- Proposer} $p$}
{\bf constants}\\
\> $A=$ set of acceptors \\
\> $\maj = \lceil(\#A+1)/2\rceil$
\\
{\bf init}
\> $\crnd\leftarrow -1$ \>\> \pcomm{current round}
\\[\smallskipamount]
{\bf on} $\tuple{\Propose,\valu}$\\
\> \pcomm{pick fresh round} \\
\> $\crnd \leftarrow \pickNextRound(\crnd)$ \\
\> $\myval \leftarrow \valu$\\
\> $P\leftarrow \emptyset$\\
\> {\bf send}  $\tuple{\Prepare,\crnd}$ {\bf to} $A$
\\[\smallskipamount]
{\bf on} $\tuple{\Promise,\rnd,\prnd,\pval}$ with $\rnd=\crnd$ \\
\> \> {\bf from} acceptor $a$\\
\> $P\leftarrow P\oplus (\prnd,\pval)$
\\[\smallskipamount]
{\bf on event} $\#P\geq \maj$\\
\> $j=\max\{\prnd \mid (\prnd,\pval)\in P\}$\\
\> {\bf if} $j\geq 0$ {\bf then}\\
\> \> $V=\{\pval \mid (j,\pval)\in P\}$\\
\> \>  \pcomm{pick value with largest $\prnd$} \\
\> \> $\myval\leftarrow \pick(V)$ \\
\> {\bf send}  $\tuple{\Accept,\crnd,\myval}$ {\bf to} $A$
\end{pseudocode}

%% file: acceptor.tex
\begin{pseudocode}[]{{Paxos --- Acceptor} $a$}
{\bf constants} $L=$ set of learners
\\
{\bf init}
\> $\crnd\leftarrow -1$ \>\> \pcomm{current round} \\
\> $\prnd\leftarrow -1$ \>\> \pcomm{previous round} \\
\> $\pval\leftarrow -1$ \>\> \pcomm{previous value}
\\[\smallskipamount]
{\bf on} $\tuple{\Prepare,\rnd}$ with $\rnd>\crnd$ \\
\>\> {\bf from} proposer $p$\\
\> $\crnd \leftarrow \rnd$\\
\> {\bf send} $\tuple{\Promise,\crnd,\prnd,\pval}$ {\bf to} $p$
\\[\smallskipamount]
{\bf on} $\tuple{\Accept,\rnd,\aval}$ with $\rnd\geq \crnd$ \\
\>\> {\bf from} proposer $p$\\
\> $\crnd\leftarrow \rnd$\\
\> $\prnd\leftarrow \rnd$\\
\> $\pval\leftarrow \aval$\\
\> {\bf send}  $\tuple{Learn,\crnd,\aval}$ {\bf to} $L$ \\[-\medskipamount]
\end{pseudocode}

%% file: learner.tex
\begin{pseudocode}{{Paxos --- Learner} $l$}
{\bf constants} \\
\> $A=$ set of acceptors \\
\> $\maj=\lceil (\#A+1)/2\rceil$ 
\smallskip\\
{\bf init}
\> $V\leftarrow \emptyset$
\\[\smallskipamount]
{\bf on} $\tuple{Learn,\rnd,\lval}$ {\bf from} acceptor $a$\\
\> $V\leftarrow V\oplus (\rnd,\lval)$
\\[\smallskipamount]
{\bf on event} $\exists m=(\rnd,\lval): \#\{m \mid \ m\in V\}\geq \maj$ \\
\> {\bf choose} $\lval$
\end{pseudocode}

%% file: spec_groove.tex
\renewcommand{\t}[1]{{\small\textsf{\bfseries #1}}}
\renewcommand{\l}[1]{{\small\textsf{#1}}}
\section{A Graph-Based Model of the Paxos Algorithm}
\label{spec}
\setgroove

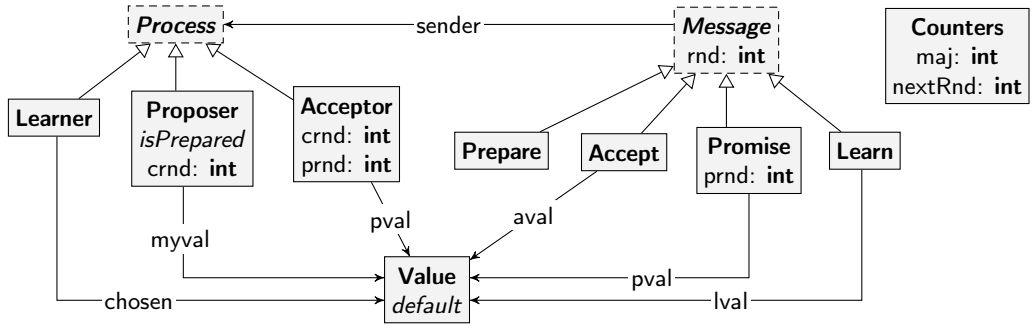
\begin{figure}[htbp] 
\centering 
\figureTextSize
  \unsetsnippet
  \input{immagini/type.tikz}%
  \setgroove

\caption{Type graph of the Paxos protocol} 
\label{fig:types} 
\end{figure}

In the graph-based model, the global states of the protocol are captured by single graphs. Each such graph is typed according to the type graph in Fig~\ref{fig:types}.

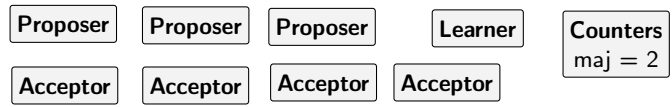
\begin{figure}[htbp] 
\figureTextSize
\centering 
  \unsetsnippet
  \input{immagini/start3-4-2.tikz}%
  \setgroove

\caption{Example initial configuration of the Paxos protocol} 
\label{fig:initial} 
\end{figure}

As the type graph shows, there are two abstract types, \t{Process} and
\t{Message}: each \t{Message} has a round number \l{rnd} during which it was
sent, and a \l{sender} (which is a \t{Process}). In addition there is a type
\t{Counters}, which will always have a singular instance that serves as a
container for the global variables \l{maj} (the bound considered to be a
majority) and \l{nextRnd} (an auxiliary variable used to dispense initial
round numbers).
There are three types of \t{Process} and four types of \t{Message}, corresponding to the roles and messages of the protocol. The arrows and attributes correspond to the local fields and variables discussed in Sect.~\ref{paxos}. In addition, the following may be noted:
\begin{itemize}[noitemsep]
\item Processes and messages have no explicit identities. This is important in
  order to ensure that symmetrical states give rise to isomorphic graphs.

\item \t{Proposer} instances have a flag \l{isPrepared}, which will be
  set when a proposer has sent a \t{Prepare} message and is ready to receive
  \t{Promise}s.

\item The values proposed and chosen by the protocol are not represented as
  integers but as nodes of type \t{Value}. The \l{default} flag on \t{Value}
  will be used to distinguished the default value with which all acceptors are
  initialized ($-1$ in the pseudocode of Fig.~\ref{pseudocode}).
\end{itemize}
Fig.~\ref{fig:initial} shows an initial configuration with three proposers, four acceptors and a majority bound of 2. The protocol is expected to be incorrect in this case, as the majority does not exceed half of the acceptors.

\begin{figure}[htbp]
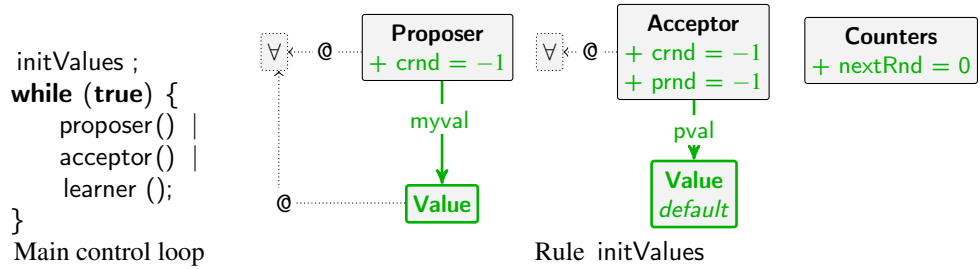
 
\figureTextSize
\centering 
\begin{tabular}{c@{{\qquad}}c}
\begin{tabular}[b]{@{}c@{}}
\inputgroove{21-26} \\[\medskipamount]
Main control loop
\end{tabular}
& \inputtikzrule{initValues}
\end{tabular}
\caption{Top-level control and initialization rule} 
\label{fig:initialization} 
\end{figure}

\paragraph{Initialization.}
The dynamics of the protocol are captured by transformation rules. In addition, the model is equipped with a control program to schedule the rules. Fig.~\ref{fig:initialization} shows the main control loop as well as the initialization rule.

The control loop specifies that the rule +initValue+ is to be invoked, followed by a perpetual choice between proposer, acceptor and learner actions, as specified by the functions +proposer()+ etc.\ (see below).

\begin{figure}[htbp] 
\figureTextSize
\centering 
  \unsetsnippet
  \input{immagini/s1.tikz}%
  \setgroove

\caption{Initial configuration of Fig.~\ref{fig:initial} after application of \l{initValues}} 
\label{fig:s1} 
\end{figure}
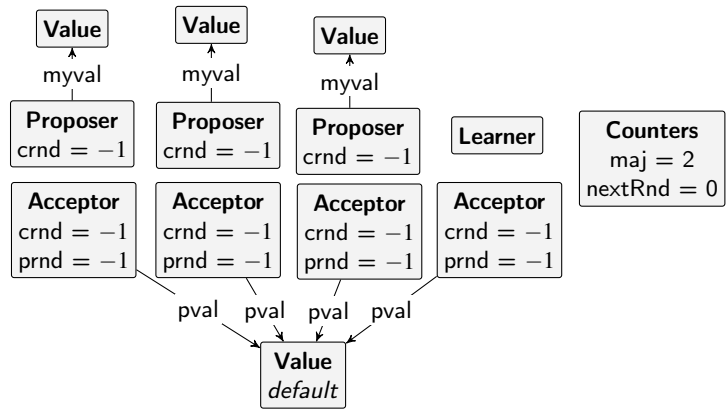

\unsetsnippet
The rule deserves clarification. The $\forall$-quantifiers cause all nodes connected with dashed \l{@}-labelled arrows to be matched as often as possible. The fat, gray \t{Value} nodes (green in a coloured view) are created as a result of the rule, as are the +-prefixed attributes in the \t{Proposer}, \t{Acceptor} and \t{Counters} nodes. For instance, applying the rule changes the graph of Fig.~\ref{fig:initial} into Fig.~\ref{fig:s1}.
\setgroove

\begin{figure}[htb]
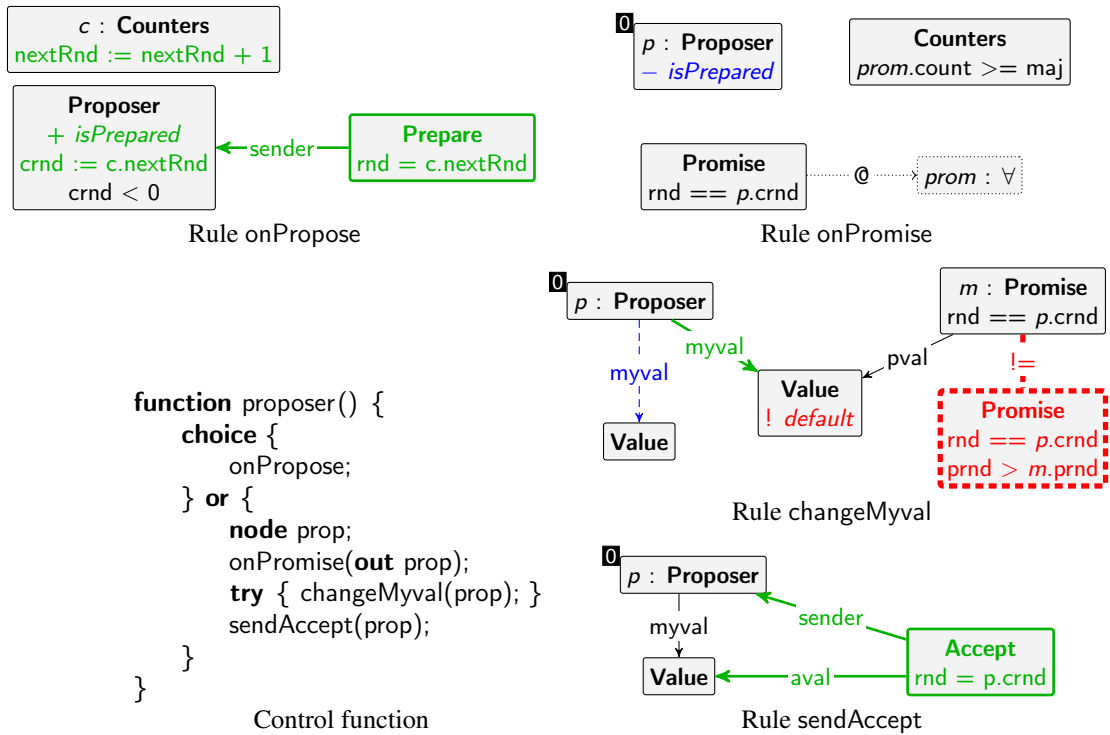

\figureTextSize
\centering
\begin{tabular}{@{}p{0.5\textwidth}@{}p{0.5\textwidth}@{}}
\inputtikzrule{onPropose}
 & \inputtikzrule{onPromise}
 \\[\medskipamount]
\hfill\begin{tabular}[b]{@{}c@{}}
\inputgroove{1-10} \\
Control function
\end{tabular}\hfill &
\begin{tabular}[b]{@{}c@{}}
\inputtikzrule{changeMyval} \\[\medskipamount]
\inputtikzrule{sendAccept}
\end{tabular}
\end{tabular}
\caption{Proposer behaviour} 
\label{fig:proposer} 
\end{figure}

\paragraph{Proposers.}
Fig.~\ref{fig:proposer} shows the control function and the rules that encompass the functionality of proposers, as specified by the pseudocode in Fig.~\ref{pseudocode}.
The control function +proposer()+ specifies a non-deterministic choice between the rule +onPropose+ on the one hand, corresponding to the ``\textbf{on} $\Propose$'' clause of Fig.~\ref{pseudocode}, and a sequence of +onPromise+, +changeMyval+ and +sendAccept+ on the other (where +try+ causes +changeMyval+ to be applied only if possible), corresponding to the ``\textbf{on} $\Promise$'' and ``\textbf{on event}'' clauses. The parameter +prop+ ensures that the rules are applied to the same proposer instance.

\begin{itemize}[noitemsep]
\item Rule +onPropose+ specifies the update of the proposer's \l{crnd} attribute and the creation of a \t{Prepare} message, under the condition that \l{crnd < 0}. Moreover, the \l{isPrepared} flag is set.

\item Rule +onPromise+  tests if the number of \t{Promise} messages with this proposer's round number exceeds the majority bound. Note that \l{prom.count}, where \l{prom} refers to the $\forall$-quantifier, stands for the number of matches of the \l{@}-connected subgraph --- in this case, just the \t{Promise} node. The \l{isPrepared} flag is a precondition for this rule and is at the same time deleted, making sure that each proposer can execute this event only once.

The adornment in the top left of the \t{Proposer} indicates that this node is a rule parameter. When the rule is applied, the value of this parameter is bound to the +prop+-variable in the control program.

\item Rule +changeMyval+ adjusts the \l{myval} field to the \l{pval} of the promise with the highest \l{prnd} value, but only if that is not the \l{default} value. (The dashed \t{Promise}-node --- red in a coloured view---  with the \l{!=}-labelled edge to the top right \t{Promise} specifies that there is \emph{no} promise with a \emph{higher} \l{prnd}.)

\item Finally, rule +sendAccept+ specifies that an \t{Accept} message is sent. Due to the scheduling in the control program, this only occurs after +changeMyval+ has had a chance to be applied.
\end{itemize}

\paragraph{Acceptors and learners.}
Fig.~\ref{fig:proposer} shows the control function and rules for the acceptor and learner roles.
\begin{figure}[htbp]
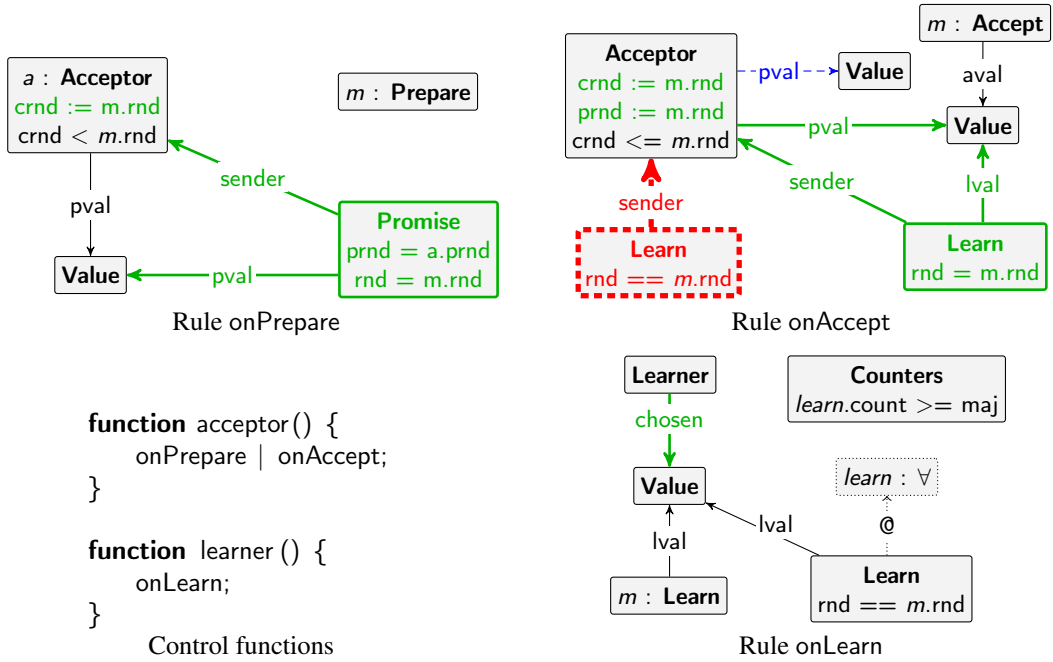
 
\figureTextSize
\centering
\begin{tabular}{p{.45\textwidth}p{.45\textwidth}}
\inputtikzrule{onPrepare}
 & \inputtikzrule{onAccept}
 \\[\medskipamount]
\centering
\begin{tabular}[b]{@{}c@{}}
\inputgroove{12-18} \\
Control functions
\end{tabular}
 & \inputtikzrule{onLearn} \hfill \\
\end{tabular}
\caption{Acceptor and learner behaviour} 
\label{fig:acceptor-learner} 
\end{figure}
The control functions merely specify a choice between rules, which in turn capture the corresponding part of the pseudocode of Fig.~\ref{pseudocode}. We will discuss the rules.
\begin{itemize}[noitemsep]
\item Rule \l{onPrepare} creates a \t{Promise} message upon discovery of a \t{Prepare} with the right \l{rnd}.
\item Rule \l{onAccept} creates a \t{Learn} message upon discovery of an \t{Accept} with the right \l{rnd}. The dashed (red) \t{Learn}-node is a negative condition ensuring that the rule is applicable at most once for any given \t{Acceptor} and \t{Accept}.
\item Rule \l{onLearn} counts the number of \t{Learn} messages with identical \l{rnd} and \l{lval} fields, in the same way as \l{onPromise} of Fig.~\ref{fig:proposer}, and chooses the corresponding \t{Value} if the count has reached the majority bound. Note that there is nothing to prevent this rule from being applied more than once; however, the same value will be chosen every time.
\end{itemize}

\begin{figure}[ht]
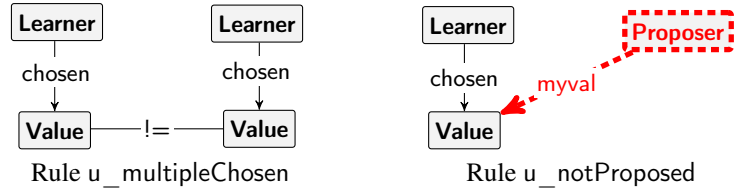

\figureTextSize
\centering
\begin{tabular}{c@{{\qquad\qquad}}c}
\inputtikzrule{u_multipleChosen} &
\inputtikzrule{u_notProposed}
\end{tabular}
\caption{Safety rules encoding the negation of the property in Def.~\ref{safety}}
\label{fig:safety} 
\end{figure}

\paragraph{Correctness.}
Based on the combination of control and rules presented above, \GROOVE can generate (and optionally visualise) the state space. Moreover, for the purpose of actually validating a model against a set of requirements, \GROOVE has built-in LTL and CTL model checkers. However, for the problem at hand, model checking is overkill, as we just want to check safety as defined in Def.~\ref{safety}.
To achieve this, it suffices to try and find graphs that are \emph{unsafe}. If that attempt fails, the protocol is correct for the initial configuration. The negated safety property is captured by the rules in Fig.~\ref{fig:safety}. Note that neither of these rules actually modifies the graph.
\begin{itemize}[noitemsep]
\item Rule~+u_multipleChosen+ tests whether two \t{Learner}s have chosen distinct \t{Value}s. The distinctness is explicitly required by the \l{!=}-edge. Due to the fact that rules may be matched non-injectively, this rule is also applicable to a graph in which a \emph{single} \t{Learner} has chosen two distinct \t{Value}s.

\item Rule~+u_notProposed+ tests whether a \t{Learner} has chosen a \t{Value} that has not been proposed by any \t{Proposer}.
\end{itemize}
\GROOVE supports a range of different exploration strategies. For this particular case it can do a depth-first search that halts as soon as a graph is found that satisfies the propositional formula +u_multipleChosen ||+  +u_notProposed+. If no such state exists, this strategy will cause the entire state space to be searched, potentially encompassing (many) millions of states.%
\unsetsnippet

%% file: immagini/type.tikz
%
\begin{tikzpicture}[scale=\tikzscale]
\node[type_node] (n4) at (5.110, -1.510) {\ml{\textbf{Promise}\\prnd: \textbf{int}}};
\node[abstract_node] (n0) at (1.300, -0.605) {\ml{\textit{\textbf{Process}}}};
\node[type_node] (n10) at (2.970, -2.370) {\ml{\textbf{Value}\\\textit{default}}};
\node[type_node] (n9) at (6.495, -0.815) {\ml{\textbf{Counters}\\maj: \textbf{int}\\nextRnd: \textbf{int}}};
\node[abstract_node] (n1) at (4.960, -0.710) {\ml{\textit{\textbf{Message}}\\rnd: \textbf{int}}};
\node[type_node] (n8) at (0.490, -1.225) {\ml{\textbf{Learner}}};
\node[type_node] (n3) at (4.280, -1.455) {\ml{\textbf{Accept}}};
\node[type_node] (n6) at (2.435, -1.335) {\ml{\textbf{Acceptor}\\crnd: \textbf{int}\\prnd: \textbf{int}}};
\node[type_node] (n7) at (1.410, -1.365) {\ml{\textbf{Proposer}\\\textit{isPrepared}\\crnd: \textbf{int}}};
\node[type_node] (n5) at (5.880, -1.445) {\ml{\textbf{Learn}}};
\node[type_node] (n2) at (3.465, -1.445) {\ml{\textbf{Prepare}}};

\path[basic_edge](n7.south -| 1.350, -2.290) -- (1.350, -2.290) -- (n10.west |- 1.350, -2.290)
node[lab] at (1.330, -2.032) {\ml{myval}};
\path[subtype_edge](n4.north -| 4.960, -0.710) --  (n1) 
;
\path[basic_edge](n5.south -| 5.860, -2.440) -- (5.860, -2.440) -- (n10.east |- 5.860, -2.440)
node[lab] at (4.990, -2.440) {\ml{lval}};
\path[subtype_edge] (n2)  --  (n1) ;
\path[subtype_edge] (n8)  --  (n0) ;
\path[basic_edge] (n3)  -- node[lab] {\ml{aval}} (n10) ;
\path[basic_edge] (n6)  -- node[lab] {\ml{pval}} (n10) ;
\path[subtype_edge] (n6)  --  (n0) ;
\path[basic_edge](n4.south -| 5.110, -2.290) -- (5.110, -2.290) -- (n10.east |- 5.110, -2.290)
node[lab] at (4.465, -2.290) {\ml{pval}};
\path[basic_edge](n1.west |- 1.300, -0.605) -- node[lab] {\ml{sender}} (n0) ;
\path[subtype_edge](n7.north -| 1.300, -0.605) --  (n0) ;
\path[subtype_edge] (n5)  --  (n1) ;
\path[basic_edge](n8.south -| 0.510, -2.440) -- (0.510, -2.440) -- (n10.west |- 0.510, -2.440)
node[lab] at (1.055, -2.440) {\ml{chosen}};
\path[subtype_edge] (n3)  --  (n1) ;
\end{tikzpicture}

%% file: immagini/start3-4-2.tikz
%
\begin{tikzpicture}[scale=\tikzscale]
\node[basic_node] (n8) at (2.615, -0.565) {\ml{\textbf{Proposer}}};
\node[basic_node] (n2) at (4.570, -0.680) {\ml{\textbf{Counters}\\maj = 2}};
\node[basic_node] (n1) at (1.775, -0.955) {\ml{\textbf{Acceptor}}};
\node[basic_node] (n6) at (3.650, -0.575) {\ml{\textbf{Learner}}};
\node[basic_node] (n0) at (0.895, -0.545) {\ml{\textbf{Proposer}}};
\node[basic_node] (n3) at (0.905, -0.955) {\ml{\textbf{Acceptor}}};
\node[basic_node] (n4) at (1.775, -0.555) {\ml{\textbf{Proposer}}};
\node[basic_node] (n5) at (2.625, -0.935) {\ml{\textbf{Acceptor}}};
\node[basic_node] (n7) at (3.445, -0.935) {\ml{\textbf{Acceptor}}};

\end{tikzpicture}

%% file: immagini/s1.tikz
%
\begin{tikzpicture}[scale=\tikzscale]
\node[basic_node] (n16) at (2.990, -0.320) {\ml{\textbf{Value}}};
\node[basic_node] (n15) at (1.145, -0.275) {\ml{\textbf{Value}}};
\node[basic_node] (n14) at (2.070, -0.250) {\ml{\textbf{Value}}};
\node[basic_node] (n13) at (2.680, -2.580) {\ml{\textbf{Value}\\\textit{default}}};
\node[basic_node] (n0) at (3.985, -1.615) {\ml{\textbf{Acceptor}\\crnd = $-1$\\prnd = $-1$}};
\node[basic_node] (n8) at (3.045, -1.025) {\ml{\textbf{Proposer}\\crnd = $-1$}};
\node[basic_node] (n4) at (1.145, -0.975) {\ml{\textbf{Proposer}\\crnd = $-1$}};
\node[basic_node] (n6) at (2.115, -1.615) {\ml{\textbf{Acceptor}\\crnd = $-1$\\prnd = $-1$}};
\node[basic_node] (n3) at (1.155, -1.615) {\ml{\textbf{Acceptor}\\crnd = $-1$\\prnd = $-1$}};
\node[basic_node] (n1) at (3.055, -1.625) {\ml{\textbf{Acceptor}\\crnd = $-1$\\prnd = $-1$}};
\node[basic_node] (n2) at (2.115, -0.985) {\ml{\textbf{Proposer}\\crnd = $-1$}};
\node[basic_node] (n7) at (5.000, -1.140) {\ml{\textbf{Counters}\\maj = 2\\nextRnd = 0}};
\node[basic_node] (n5) at (3.970, -0.985) {\ml{\textbf{Learner}}};

\path[basic_edge](n8.north -| 2.990, -0.320) -- node[lab] {\ml{myval}} (n16) ;
\path[basic_edge](n4.north -| 1.145, -0.275) -- node[lab] {\ml{myval}} (n15) ;
\path[basic_edge](n2.north -| 2.070, -0.250) -- node[lab] {\ml{myval}} (n14) ;
\path[basic_edge] (n0)  -- node[lab] {\ml{pval}} (n13) ;
\path[basic_edge] (n6)  -- node[lab] {\ml{pval}} (n13) ;
\path[basic_edge] (n3)  -- node[lab] {\ml{pval}} (n13) ;
\path[basic_edge] (n1)  -- node[lab] {\ml{pval}} (n13) ;
\end{tikzpicture}

%% file: spec_spin.tex
\section{A Vector-Based Model of the Paxos Algorithm}
\setpromela

We now present a formal specification of Paxos in \PROMELA, the input language of the model checker \SPIN. A \PROMELA specification consists of a number of processes communicating through shared channels. The processes are obtained by instantiating so-called +proctype+ templates. We call this type of model \emph{vector-based} because of the way the states are encoded during state space exploration, namely as vectors of values (representing local variables and channel contents).

\paragraph{Initialization.}
For Paxos, we define three +proctype+s, corresponding to the roles of the protocol. For instance, the initial configuration shown for \GROOVE in Fig~\ref{fig:initial} is specified in \PROMELA as
\inputpromela{80-84}
We will discuss the template definitions and their parameters below. The channels and majority bound are defined as follows:
\inputpromela{1-6}
Every channel conveys messages of one of the four types in the protocol. Any process can access any of the channels for sending or receiving. For the purpose of this protocol we always use +!!+ to insert messages in channels in lexicographic order and +??+ to select the first matching message from them; this effectively turns the channels into multisets of messages while keeping channels in canonical form (ordered lists).
\begin{figure}[ht]
\figureTextSize
\centering
\begin{tabular}{@{}c@{{\quad}}c@{}}
\hline
\begin{tabular}[t]{@{}l@{}}\inputpromela{86-120}\end{tabular} &
\begin{tabular}[t]{@{}l@{}}\inputpromela{40-63}\end{tabular} \\
\hline
\multicolumn{2}{c}{\inputpromela{8-13}} \\
\hline
\end{tabular}
\caption{Proposer and acceptor templates, with inlined broadcast primitives.}
\label{proposer}
\end{figure}

\paragraph{Templates.}
Fig.~\ref{proposer} shows the proposer and acceptor templates. Proposers are identified by a unique value passed in as the first parameter, +crnd+. The other parameter, +myval+, is the proposed value. The template starts with a +bprepare+ invocation, which causes a broadcast of +prepare+ messages.
This is followed by a non-deterministic +do+--+od+ loop which is exited once an +accept+ message has been broadcast. In this loop, in an atomic step, first the messages in the +promise+ buffer with the right +rnd+ parameter are counted by iterating over all the messages (using a temporary artificial message with non-existent round number as marker); then, if the count exceeds the majority, the +accept+ is broadcast.

The +acceptor+ template of Fig.~\ref{proposer} is very similar to the pseudocode of Fig.~\ref{pseudocode}. The +id+ parameter is required to address individual acceptors in +prepare+ and +accept+ messages.
Finally, the learner template is defined in Fig. \ref{learner}. The counters +mcount[rnd]+ keep track of the number of received +learn+ messages associated to round +rnd+.

\begin{figure}[ht]
\figureTextSize
\centering
\begin{tabular}{c}
\hline
\inputpromela{65-78} \\
\hline
\end{tabular}
\caption{Learner process template.}
\label{learner}
\end{figure}

\paragraph{Modelling choices.}
In order to keep the state space size as small as possible, we have taken the following measures in the \PROMELA model:
\begin{itemize}[noitemsep]
\item All communication uses the ordered send and unordered receive operators +!!+ and +??+.
\item Sequences of statements are wrapped in +d_step+ and +atomic+ blocks wherever possible. Their execution is thus represented by a single transition, not interrupted by other processes (unless a statement blocks inside an +atomic+ statement).
\item The +proposer+ uses a ``quorum transition'' scheme for atomically counting the relevant messages, while avoiding a state change in the +promise+ buffer.
\item Local variables are reset to their initial values when they are no longer needed.
\item Message count variables are no longer increased after they reach the majority bound.
\end{itemize}

 
\unsetsnippet

%% file: analysis.tex
\section{Analysis}
\input{groove_experiments}
\input{spin_experiments_revised}

%% file: groove_experiments.tex
\textbf{\GROOVE experimental results.}
Table~\ref{groove-results} shows the outcome of running \GROOVE on the graph-based model with different start states. These results were obtained using an Intel i7--2600 (64~bits) CPU at 3.4 GHz under Windows 7, running Java~6 with 8~GB of memory. 

Each row reports the state count and time used for state space exploration with a given number of proposers (first column) and acceptors (second column), for a majority bound that is too low (columns 3--5), respectively just high enough (columns 6--8) for the protocol to be correct. In all cases where the bound is too low, a violation of the safety properties (Def.~\ref{safety}) was found well before the full state space was explored; in contrast, with the bound high enough, no violations were found and hence all reachable states were exhaustively enumerated. The number of states in the second case is typically several orders of magnitude higher than in the first case.
Every next step not reported in the table (a higher number of proposers for the same number of acceptors, or vice versa) causes the full state space exploration to run out of memory.

\begin{table}[htb]
\figureTextSize
\centering
\begin{tabular}[t]{|c@{\;}c|c@{\;}r@{\;}r|c@{\;}r@{\;\;}r|}
\hline\hline
 &
 & \multicolumn{3}{l|}{\bf Low majority (w.\ iso)}
 & \multicolumn{3}{l|}{\bf High majority (with iso)}
 \\
\bf Proc & \bf Acc
 & \bf Maj & \bf States & \bf Time (ms)
 & \bf Maj & \bf States & \bf Time (ms)
 \\
\hline
\input{tests/iso/table.tex}
\hline
\end{tabular}%
\begin{tabular}[t]{c@{\;}rr|}
\hline\hline
\multicolumn{3}{l|}{\bf High majority (no iso)} \\
\bf Maj & \bf States & \bf Reduct \\\hline
\input{tests/compare.tex}
 \hline
\end{tabular}
\caption{\GROOVE model checking results, with and without isomorphism reduction}
\label{groove-results}
\end{table}

\begin{figure}[htb]
\figureTextSize
\centering
\includegraphics[scale=.53]{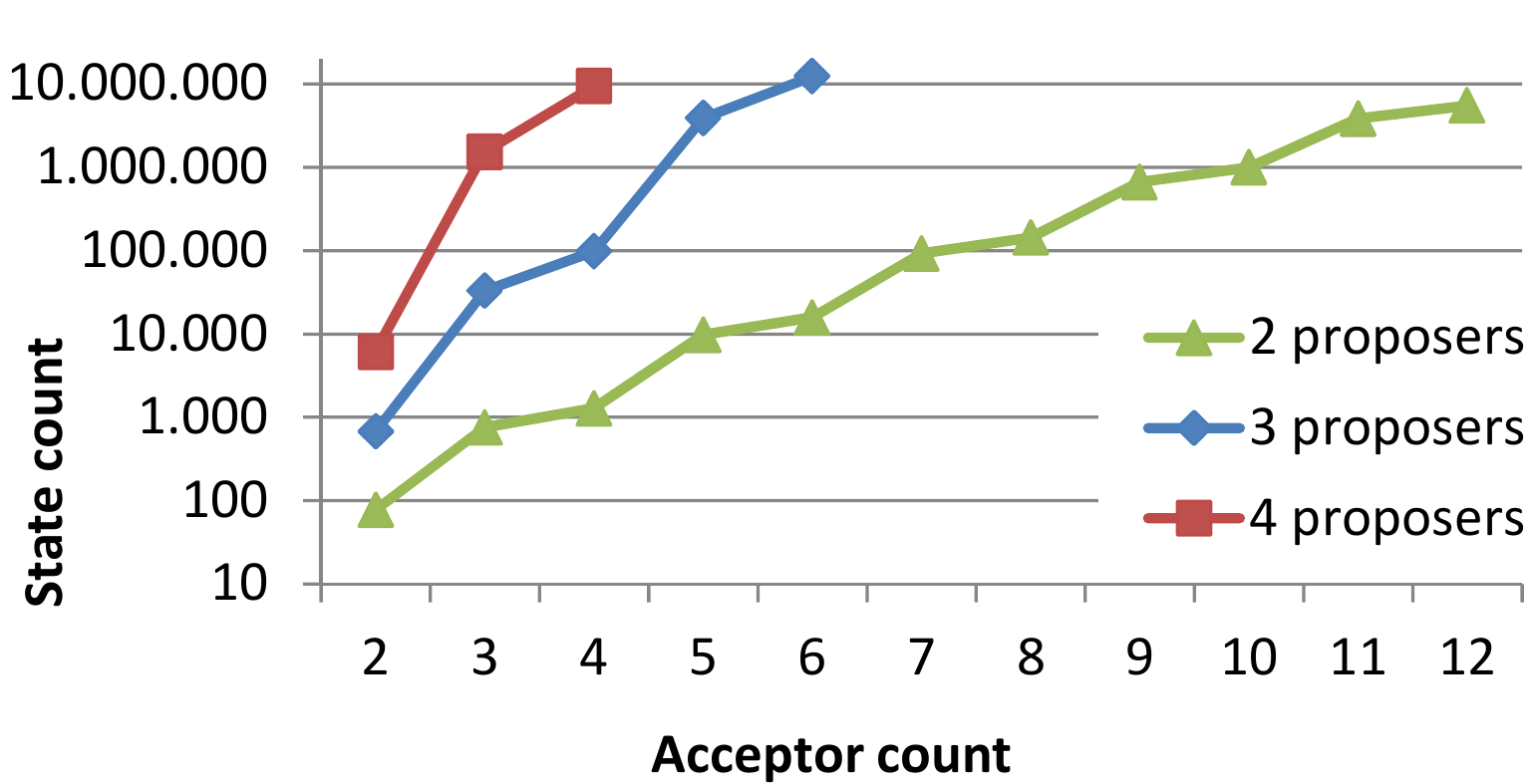}%
\includegraphics[scale=.44]{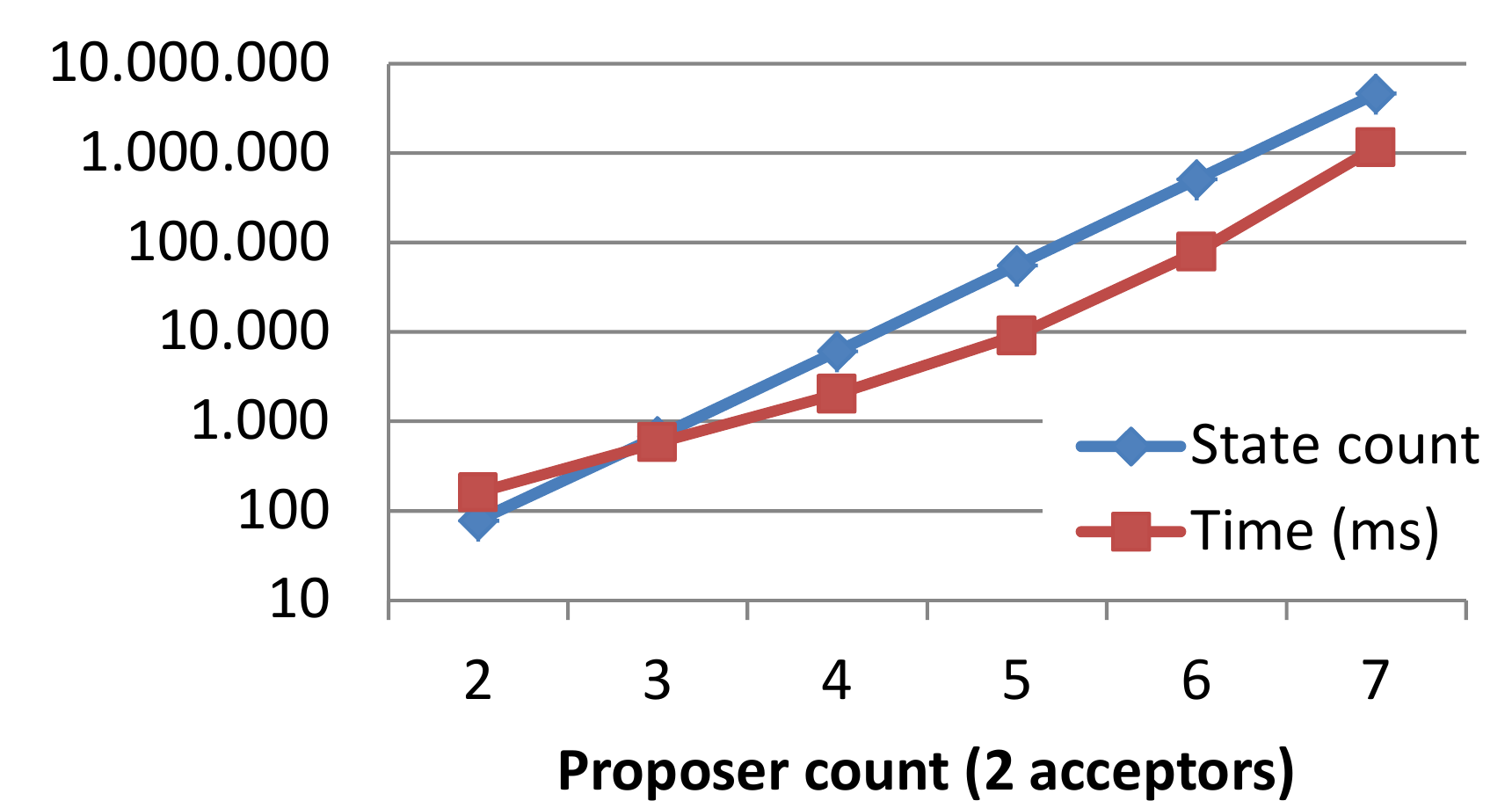} 
\caption{Graphs for the results of Table~\ref{groove-results}}
\label{groove-result-graphs}
\end{figure}

Figure~\ref{groove-result-graphs} shows two graphs derived from the results in Table~\ref{groove-results}.
The first graph shows how the state space size (for full exploration) scales with the number of acceptors, for 2, 3 and~4 proposers. We call attention to the following phenomena:
\begin{itemize}[noitemsep]
\item The vertical axis has a logarithmic scale; clearly the state space grows exponentially with the number of acceptors. Moreover, the growth much is accelerated for larger numbers of proposers, from less then 1 order of magnitude for each next acceptor (2 proposers) to around 2 orders of magnitude (4 proposers)

\item The increase has a slight sawtooth shape, due to the fact that the majority bound does not increase with every next acceptor, but with every \emph{second} additional acceptor.
\end{itemize}
The second graph shows the state space size and running time for an increasing number of proposers and a fixed number of 2 acceptors. The following can be remarked:
\begin{itemize}[noitemsep]
\item The increase is again exponential; in fact, the exponent is larger than for the case of 2 proposers and increasing number of acceptors. This can be explained by the fact that, in contrast to the acceptors, the proposers essentially do \emph{not} engender symmetry, since they are assigned identities through the $\crnd$ attribute.

\item The time required by the analysis essentially keeps step with the state space size, being slightly worse for smaller problem sizes due to initialization effects and for larger problem sizes due to garbage collection.
\end{itemize}
Finally, we want to draw attention to the tool performance, in terms of number of states generated per second. This fluctuates for different problem sizes, from under 400 to over 6000. The symmetries in the pool of acceptors cause the analysis to be slower for $p$ proposers and $a$ acceptors than for $a$ proposers and $p$ acceptors, as the isomorphism checking necessary to detect these symmetries is computationally expensive (see \cite{GROOVE-iso} for an extensive discussion of the principles behind the symmetry reduction of \GROOVE). In fact, profiling shows that for $(p,a)=(2,11)$, which has the lowest rate of generated states/s, over 75\% of the total computation time is spent in isomorphism checking.

\noindent\textbf{Isomorphism reduction.}
The last two columns of Table~\ref{groove-results} also show the reduction due to isomorphism checking, for those problem instances for which \GROOVE was able to compute the state space without reduction. Clearly, the gain is enormous, and easily justifies the time spent in computing isomorphism. The same effect was reported, more extensively, in \cite{Crouzen+}.



%% file: tests/iso/table.tex
2 & 2 & 1 & 45 & 110 & 2 & 78 & 160 \\
2 & 3 & 1 & 55 & 150 & 2 & 757 & 790 \\
2 & 4 & 2 & 174 & 360 & 3 & 1,279 & 1,460 \\
2 & 5 & 2 & 222 & 550 & 3 & 9,729 & 6,970 \\
2 & 6 & 3 & 723 & 1,820 & 4 & 15,783 & 11,521 \\
2 & 7 & 3 & 911 & 3,100 & 4 & 92,289 & 69,316 \\
2 & 8 & 4 & 2,574 & 4,960 & 5 & 143,376 & 131,028 \\
2 & 9 & 4 & 3,154 & 4,982 & 5 & 665,564 & 844,890 \\
2 & 10 & 5 & 7,729 & 10,581 & 6 & 992,044 & 1,529,461 \\
2 & 11 & 5 & 9,233 & 29,963 & 6 & 3,820,671 & 7,698,559 \\
2 & 12 & 6 & 20,192 & 40,804 & 7 & 5,491,406 & 14,794,452 \\
\hline
3 & 2 & 1 & 51 & 120 & 2 & 677 & 581 \\
3 & 3 & 1 & 62 & 160 & 2 & 32,899 & 7,032 \\
3 & 4 & 2 & 610 & 842 & 3 & 98,330 & 25,983 \\
3 & 5 & 2 & 407 & 900 & 3 & 3,880,277 & 6,681,550 \\
3 & 6 & 3 & 1,481 & 3,411 & 4 & 12,247,549 & 8,771,175 \\
\hline
4 & 2 & 1 & 58 & 140 & 2 & 6,082 & 2,020 \\
4 & 3 & 1 & 246 & 350 & 2 & 1,523,338 & 940,095 \\
4 & 4 & 2 & 1,258 & 1,710 & 3 & 9,337,923 & 4,523,411 \\
\hline
5 & 2 & 1 & 66 & 150 & 2 & 55,420 & 9,131 \\
\hline
6 & 2 & 1 & 75 & 170 & 2 & 506,370 & 77,888 \\
\hline
7 & 2 & 1 & 85 & 190 & 2 & 4,607,455 & 1,154,561 \\

%% file: tests/compare.tex
2 & 224 & 65.18\% \\
2 & 6,882 & 89.00\% \\
3 & 31,256 & 95.91\% \\
3 & 1,178,114 & 99.17\% \\
4 & -- & -- \\
4 & -- & -- \\
5 & -- & -- \\
5 & -- & -- \\
6 & -- & -- \\
6 & -- & -- \\
7 & -- & -- \\
\hline
2 & 6,674 & 89.86\% \\
2 & 1,079,582 & 96.95\% \\
3 & -- & -- \\
3 & -- & -- \\
4 & -- & -- \\
\hline
2 & 260,910 & 97.67\% \\
2 & -- & -- \\
3 & -- & -- \\
\hline
2 & 12,743,315 & 99.57\% \\
\hline
2 & -- & -- \\
\hline
2 & -- & -- \\
\hline

%% file: spin_experiments_revised.tex
\noindent\textbf{\SPIN experimental results.}
To compare graph- and vector-based representations, we have applied the Spin
model checker to the Paxos Promela model presented above.  The experiments in
Table \ref{spin-results} are obtained on an i7 processor with 6GB of RAM
running jSpin (spin 6.2.5 May 2013) under Windows 7 with default option for memory
management.
\begin{table}[htb]
\figureTextSize
\centering
\scalebox{0.93}{
\begin{tabular}{|c@{~~~}c|c@{}r@{~~~}r@{~~~}rr@{~~~}|c@{}r@{~~~}r@{~~~}r@{~~~}r@{~~~}r|}
\hline\hline
 &
 & \multicolumn{5}{l|}{\bf Low majority}
 & \multicolumn{6}{l|}{\bf High majority}
 \\
\bf P& \bf A
 & \bf Maj & \bf States & \bf States$^*$ & \bf Time & \bf Time$^*$
 & \bf Maj & \bf States & \bf States$^*$ & \bf hf & \bf Time & \bf Time$^*$
 \\
\hline
2  & 2  &  1   &    338      & 338      & 6       & 5      & 2    &     620   &     62 0    &     &     6   & 8\\
2  & 3  &  1   &    1,671    & 1,671    & 22      & 22     & 2  &  29,352   &   29,352      &     &   277   & 272\\
2  & 4  &  2   &    40,000   & 40,000   & 3,339   &  3,270 & 3 & 269,419   &  269,419       &     & 3,5     & 3,510\\
2  & 5  &  2   &   226,306   & 387,982  & 4,470   & 4,480  & 3 &    --     & 20,808,669     & 6,5 &   --    &    229,000\\
2  & 6  &  3   &  4,374,568  & 4,374,568 & 118,000 &  62,900  & 4 &    --     & 60,221,751  & 2,2 &   -- & 72,950,000\\
%
%
\hline
3  & 2  &  1   &    3,533  & 3,533        &  31 & 26          & 2    &   19,536 & 19,536       &    & 197 & 114 \\
3  & 3  &  1   &    37,868 & 37,868       & 643 & 362         & 2    &   4,998,934 & 4,997,226 & 26 & 71,700 & 43,700\\
3  & 4  &  2   &  2,712,145 &   2,711,977 & 68,400 & 37,000   & 3    &    -- &  59,296,034      & 2,3 & -- & 706,000\\
\hline  
4  & 2  &  1   &    38,628   & 38,628      &   9,050   & 5,230       & 2  &  617,129  & 617,129 & & 8,500    & 5,600 \\
4  & 3  &  1   &   870,245 & 870,245    &  21,600   & 12,100     & 2  &  -- & 26,245,402  & 5,1 & -- & 339,000\\
4  & 4  &  2   &   --  & 65,115,933 &  -- & 1,290,000  & 3  &  -- & 70,809,409 & 1,9& -- & 1,350,000 \\
\hline
\end{tabular}
}
\caption{\SPIN model checking results (time is in ms): 
-- indicates an out of memory error; $^*$ indicates results obtained via bit-hashing: hf indicates the hash factor (it indicates a high level of coverage when $>$100).}
\label{spin-results}
\end{table}
Without use of underapproximated search, with 2 proposers Spin is capable of finding violations up to 5 acceptors,
whereas it can prove safety up to 4 acceptors.
With 3 proposers,  violations cannot be detected with more than 3 acceptors, and safety cannot be proved for more than
2 acceptors. A similar result is obtained for 4 proposers.
By applying underapproximation heuristics like bit-hashing, Spin can deal with larger configurations while still detecting violations with low majorities, as shown by the $*$-marked columns in Table \ref{spin-results}.

%% file: conclusions.tex
\begin{figure}[htb]
\figureTextSize
\centering
\includegraphics[scale=.60]{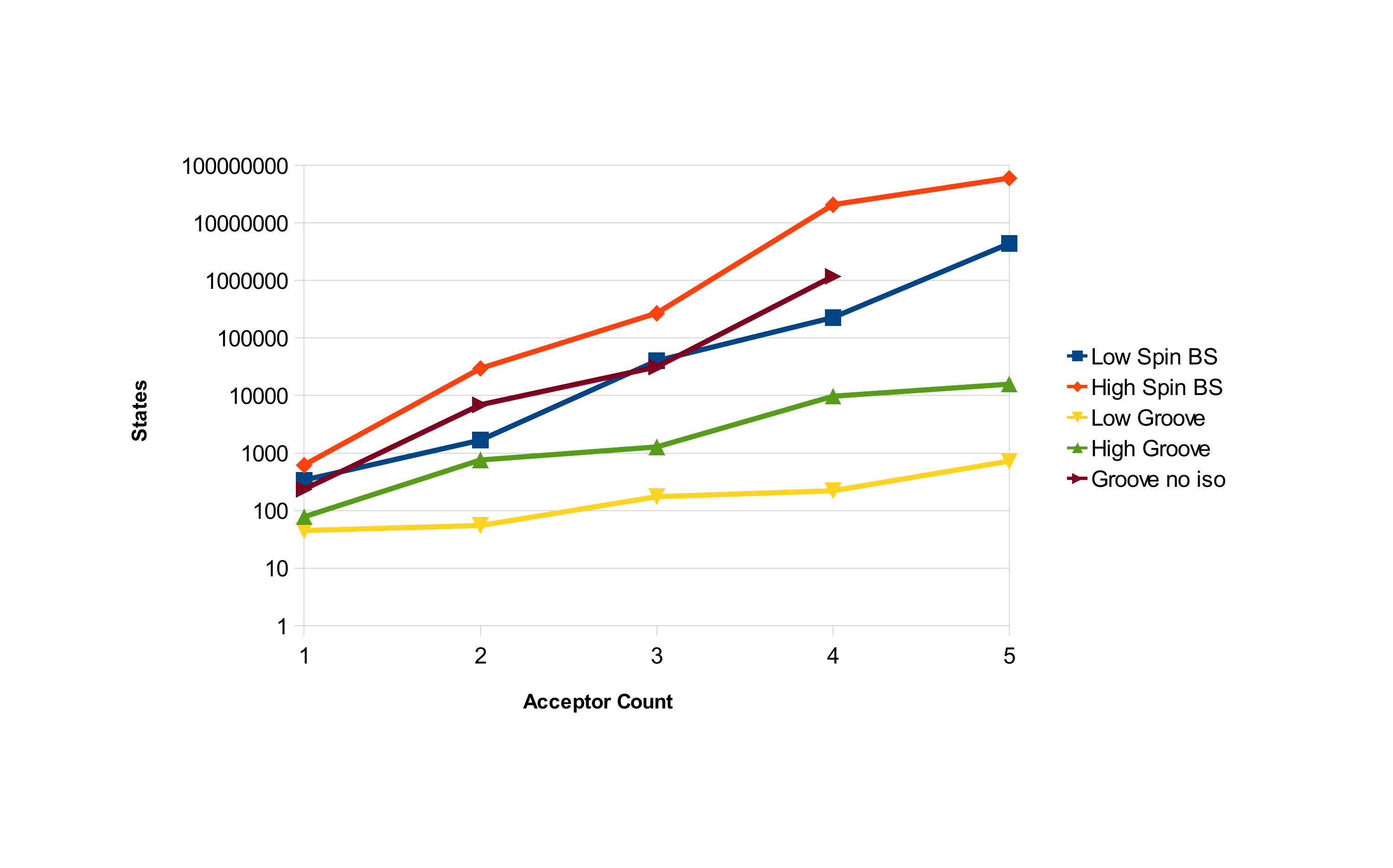}%
\caption{\GROOVE versus \SPIN state space sizes, 2 proposers}
\label{groove-spin-graph}
\end{figure}
\section{Conclusions}
\label{conclusions}
In this paper we have presented a declarative model of the Paxos consensus algorithm obtained via graph transformation rules as those used as input language for the Groove simulator and model checker.
The use of extended graph transformation rules, e.g., with negative conditions and nested quantification, allows to naturally compile pseudo-code in a declarative specification in Groove.
Furthermore, when compared to analysis with more traditional verification tools like Spin, experimental results with the Groove model checker show an impressive reduction of the state-space obtained via symmetry reductions based on graph isomorphism.
These reductions fully exploit the underlying graph-representation of the configurations in which a key point is to use anonymous nodes to denote values and identifiers (without need of introducing integers or other enumerative types).
For two proposers, Figure~\ref{groove-spin-graph} shows the difference, in logarithmic scale, between Groove with iso-check and Spin with bitstate hashing. 

The considered case-study and experimental comparison show that, well engineered graph-based search engines like Groove can compete with vector-based enumerative engines on non trivial protocol case-studies like Paxos.
Comparisons with symbolic model checkers on this kind of protocols and other examples of distributed algorithms seems an interesting research direction to understand the limit of declarative model checker tools like Groove.  